\documentclass[prb,twocolumn,showpacs]{revtex4}
\usepackage{epsfig,color}

\begin{document}
\newcommand{\ltwid}{\mathrel{\raise.3ex\hbox{$<$\kern-.75em\lower1ex\hbox{$\sim$}}}}
\newcommand{\gtwid}{\mathrel{\raise.3ex\hbox{$>$\kern-.75em\lower1ex\hbox{$\sim$}}}}
\newcommand{\bra}{\langle}
\newcommand{\ket}{\rangle}
\newcommand{\sill}{\psi}
\newcommand{\trace}{{\rm Tr}}
\newcommand{\ntilde}{\tilde{n}}
\newcommand{\stilde}{\tilde{s}}
\newcommand{\atilde}{\tilde{\alpha}}
\newcommand{\greg}[1]{\textcolor{red}{#1}}
\newcommand{\adrian}[1]{\textcolor{blue}{#1}}
\newcommand{\question}[1]{\textcolor{green}{#1}}

\title{Spectral properties of a spin-incoherent Luttinger Liquid}

\author{Adrian E. Feiguin}
\affiliation{Department of Physics and Astronomy, University of Wyoming, Laramie, Wyoming 82071, USA}

\author{Gregory A. Fiete}
\affiliation{Department of Physics, The University of Texas at Austin, Austin, Texas 78712, USA}

\date{\today}

\begin{abstract}
We present time-dependent density matrix renormalization group (DMRG) results for strongly interacting one dimensional fermionic systems at finite temperature.   When interactions are strong the characteristic spin energy can be greatly suppressed relative to the characteristic charge energy, allowing for the possibility of spin-incoherent Luttinger liquid physics when the temperature is high compared to the spin energy, but small compared to the charge energy.  Using DMRG we compute the spectral properties of the $t-J$ model at arbitrary temperatures with respect to both spin and charge energies.  We study the full crossover from the Luttinger liquid regime to the spin-incoherent regime,
focusing on small $J/t$, where the signatures of spin-incoherent behavior are more manifest. Our method allows us to access the analytically intractable regime where temperature is of the order of the spin energy, $T\sim J$.
 Our results should be helpful in the interpretation of experiments that may be in the crossover regime, $T\sim J$, and apply to one-dimensional cold atomic gases where finite-temperature effects are appreciable.  The technique may also be used to guide the development of analytical approximations for the crossover regime.

\end{abstract}

\maketitle

\section{Introduction} 
A remarkable aspect of one-dimensional interacting electron systems (we will use one-dimensional electrons as a concrete example throughout the paper, but our results immediately generalize to any system with an internal ``spin" degree of freedom, {\it e.g.} cold atomic gases)  is that a perturbative treatment of the interactions about the non-interacting limit, and thus Fermi liquid theory, fails. This is a consequence of the pervasive nesting taking place at all densities and polarizations, due to the simple fact that the Fermi surface reduces to two Fermi points.  The central result to emerge in one dimension is that, for many realistic parameters, the low-energy physics is gapless and described by a universal low-energy theory called ``Luttinger liquid"  (LL) theory.\cite{GiamarchiBook,GogolinBook,Haldane1981}  According to LL theory, there are no electron-like quasi-particles analogous to those found in Fermi liquid theory (which describes interacting electrons in three dimensions).  Instead, the low energy physics is dominated by bosonic collective excitations.  LL theory also states that for finite interactions there will be a spin-charge separation with distinct collective spin and charge excitations that each have their own characteristic velocity and Hamiltonian.  The spectral properties of the LL are very different from a Fermi liquid, but have been computed and are known.\cite{Voit1993,Meden1992,Iucci2007}

A particularly good realization (low disorder) of one-dimensional electrons is found in high mobility semiconductor heterostructures of the type often used to study the fractional quantum Hall effect.  Related systems were used recently to establish the presence of  LL physics in quantum wires.\cite{Auslaender2002,Auslaender2005}  While previous carbon nanotube experiments \cite{Yao1999,Bockrath1999} demonstrating a power-law form of the tunneling density of states were correctly interpreted as an indication of LL behavior, they did not unambiguously establish its existence because they did not probe the full spectral function of the system due to the local tunneling of electrons (which does not allow momentum resolution).  The key feature of the semiconductor heterostructure devices is that parallel wires can be fabricated and momentum-resolved tunneling experiments performed.  It is the momentum resolution that allowed the dynamical properties of the wires to be measured and LL behavior to be unambiguously observed.\cite{Auslaender2002,Auslaender2005}

However, these experiments also showed a distinct set of behaviors when the temperature was estimated to be of the order or much larger than the characteristic spin energy.\cite{Steinberg2006}  In this regime, LL theory is not expected to hold, but rather a separate theory describing ``spin-incoherent" electrons takes over. \cite{Fiete2007b}  It turns out this ``spin-incoherent Luttinger liquid" (SILL) has many {\em more} universal properties than the LL, \cite{Fiete2007b} but its conclusive demonstration in experiment is not yet universally agreed upon.\cite{Halperin2007}  Part of the challenge is that for realistic parameters many one-dimensional systems fall in the crossover regime between LL and SILL,  greatly complicating the interpretation of the experiments.\cite{Hew2008,Deshpande2008}  This crossover regime is not easily or accurately handled by existing analytical methods, so a numerical approach is required.  

In this work, we describe a technique well suited to this challenge and compute several quantities that can be directly compared to experiment.  The qualitative agreement with existing experiments in Ref.[\onlinecite{Steinberg2006}] provides further evidence that the spin-incoherent regime has indeed been reached.  At present, there are no other numerical methods that have been demonstrated to accurately access the parameter regimes and system sizes we study here.  A strength of our density matrix renormalization group (DMRG)-based calculation is that it free of the artifacts introduced by statistical sampling, such as in quantum Monte Carlo.  Our calculations start deep within the spin-incoherent regime and approach the crossover regime ``from above" (i.e. from temperatures above than the crossover temperature), complementing existing analytical methods that attempt to approach the crossover regime ``from below" using LL theory. Most importantly, the method accurately captures the crossover regime allowing its properties to be revealed.  Such information can be crucial in the proper interpretation of experiments in the strongly interacting regime (which are always at finite temperature) and can be used as an aid in the development of approximate analytical methods to describe this regime. 

\begin{centering}
\begin{figure}
\epsfig {file=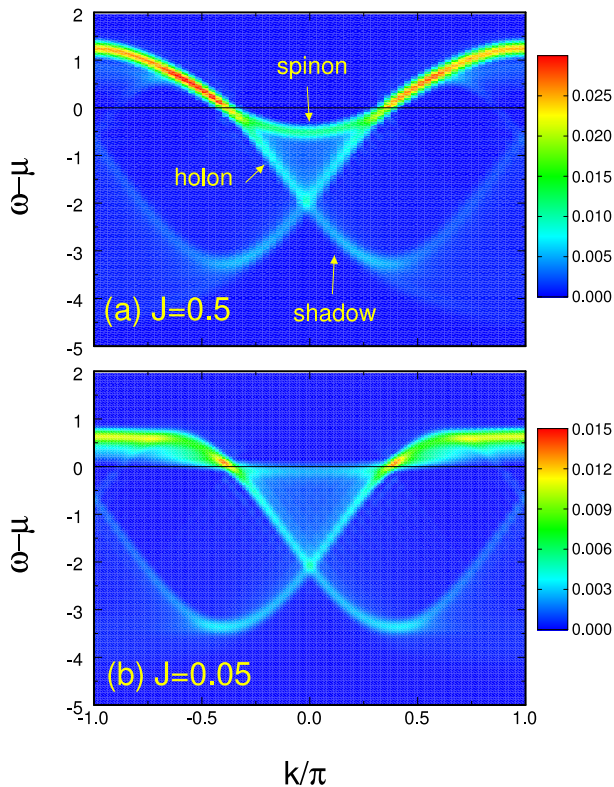,width=80mm}
\caption{ 
Momentum resolved spectrum at zero temperature of a $t-J$ chain of length $L=64$, with $N=48$ particles, and (a) $J=0.5$ (b) $J=0.05$ (in units of $t$), obtained with time-dependent DMRG. Negative frequencies correspond to the photoemission spectrum obtained by removing a fermion, while positive values correspond to inverse-photoemission. The spinon bands have a weak dispersion of width $\sim J$. Holon, and corresponding shadow bands, are clearly visible.  Frequencies are measured in units of $t$. 
} 
\label{fig1}
\end{figure}
\end{centering}

\section{Spin-incoherent Hubbard chain}
To be concrete, we study the one-dimensional Hubbard model:
\begin{equation}
H=-t \sum\limits^L_{i=1,\sigma} \left(c^\dagger_{i\sigma} c_{i+1\sigma}+\mathrm{h.c.}\right)
+ U \sum\limits^L_{i=1} n_{i\uparrow} n_{i\downarrow},
\label{hubbard}
\end{equation}
where $c^\dagger_{i\sigma}$ creates an electron of spin $\sigma$ on the
$i^{\rm th}$ site along a chain of length $L$. The hopping parameter of the Hubbard chain is $t$,
the onsite interaction energy is $U$,  and we take the inter-atomic distance as 
unity.
In the limit of large repulsive $U$, we can equivalently consider the 
$t-J$ model, defined as
\begin{eqnarray}
H_{t-J} &=& -t \sum\limits^L_{i=1,\sigma} \left(c^\dagger_{i\sigma} c_{i+1\sigma}+\mathrm{h.c.}\right)  \\
&+& J \sum\limits^L_{i=1} \left(\vec{S}_i \cdot \vec{S}_{i+1} - \frac{1}{4} n_i n_{i+1} \right),
\label{tj}
\end{eqnarray}
where the constraint forbidding double-occupancy has been imposed.
The natural excitations of this model are charge and spin collective modes 
(holons and spinons, respectively) with different velocities that depend on the 
ratio $U/t$, or $J/t$.
In the $U\to \infty$, $J\to0$ limit, the ground state  
factorizes into the 
product of a fermionic wave function $|\phi\rangle$, and a spin wave function 
$|\chi\rangle$ \cite{Ogata1990} 
\begin{equation}
|\mathrm{g.s.}\rangle=|\phi\rangle\otimes |\chi\rangle.
\label{gs}
\end{equation}
The first piece, $|\phi\rangle$, describes the charge degrees of freedom, and is 
simply the ground state of a spinless non-interacting tight-binding Hamiltonian. 
At finite $U$, the spins are governed by a Heisenberg interaction
\begin{equation}
H_s = J\sum_{i}\vec{S}_{i} \cdot \vec{S}_{i+1},
\label{heisenberg}
\end{equation}
where $J$ depends on the charge wave-function and is proportional to $4t^2/U$.\cite{Xiang1992}
In the $U\to \infty$, $J\to 0$ limit, the spin states are degenerate and the dispersion 
is just a non-interacting band $\epsilon(k)=-2t\cos(k)$, but any finite interaction will lift this degeneracy and give the spin degree of freedom some dispersion.  The factorized wavefunction approach has been used in a number of key earlier studies\cite{Ogata1990,Xiang1992,Penc1996,Favand1997} in the strongly interacting regime, but {\em we do not make such an approximation here}.

In Fig.\ref{fig1} we show the momentum resolved spectrum of a chain with $L=64$ sites and $N=48$ particles, 
for values of $J=0.5$ and $J=0.05$ (all quantities are in units of the hopping), obtained with time-dependent DMRG at zero temperature.\cite{White2004a, Daley2004} The spectrum is clearly gapless, displaying a weakly dispersive spinon band of width $\sim J$, and broad holon bands of width $\sim 4t$. Our results agree in the $U \to \infty$ limit with the dispersion calculated in Refs.[\onlinecite{Penc1996,Favand1997}] using the factorized wave function, Eq. (\ref{gs}), and also with the exact diagonalization results for the $t-J$ model in Ref.[\onlinecite{Favand1997}], at the same density and similar value of parameters. 
For small values of $J$ we see an almost non-dispersive spinon band. In this case, a spin-incoherent behavior would be observed at a finite temperature larger than 
the characteristic spin energy scale, but much smaller than the Fermi energy $J 
\ll T \ll E_F\sim t$. When these conditions are realized, the spins are totally 
incoherent, effectively at infinite temperature, while the charge sector remains very close to the ground state.

\begin{centering}
\begin{figure}
\epsfig {file=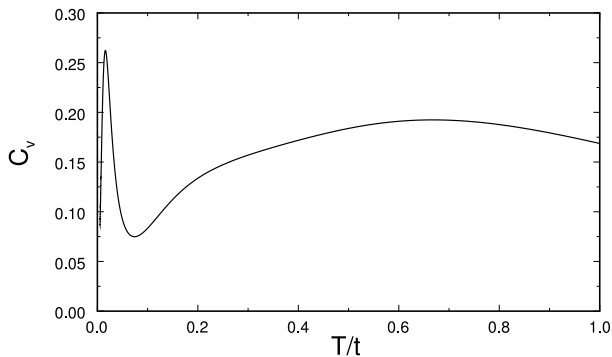,width=80mm,angle=-00}
\caption{ 
Specific heat of a $t-J$ chain of length $L=32$ and $N=24$ fermions, calculated with time-dependent DMRG. Temperature is in units of the hopping $t$.
} 
\label{fig2}
\end{figure}
\end{centering}

\section{Method}
The key idea behind our calculation is thermo field dynamics.\cite{Takahashi1975,UmezawaBook,Matsumoto1986,Suzuki1985,Suzuki1985b,Barnett1988,Barnett1985} This construction allows one to 
represent a mixed state of a quantum system as a pure state in an enlarged 
Hilbert space.
Consider the energy eigenstates of the system in
question $\{n\}$, described by a Hamiltonian $H$, and introduce an auxiliary set 
of fictitious states $\{\tilde{n}\}$ in one-to-one correspondence with $\{n\}$.
We can then define the unnormalized pure quantum state,
\begin{equation}
| \psi(\beta) \rangle = e^{-\beta H/2}| \psi(0) \rangle = \sum_n
e^{-\beta E_n/2} |n \tilde{n}\rangle \label{thermoa}
\end{equation}
where $\tilde{n}$ is a copy of $n$ in the auxiliary Hilbert space, $\beta=1/T$ is the 
inverse temperature, and $|\psi(0)\rangle=\sum_n{|n\tilde{n}\rangle}$ is our
thermal vacuum.
Then the exact thermodynamic average of an operator $\hat O$ (acting
only on the real states), is given by 
\begin{equation}
\langle \hat O \rangle = Z(\beta)^{-1} \langle \sill(\beta) | \hat O | \sill(\beta) \rangle,
\label{average}
\end{equation}
Where the partition function is the norm of the thermal state $Z(\beta)=\langle 
\psi(\beta) | \psi(\beta) \rangle$. 
We can clearly see how the calculation of a thermodynamic average reduces to 
calculating a conventional expectation value of an operator in a pure quantum 
state, at the price of working in a larger Hilbert space.

\begin{centering}
\begin{figure}
\epsfig {file=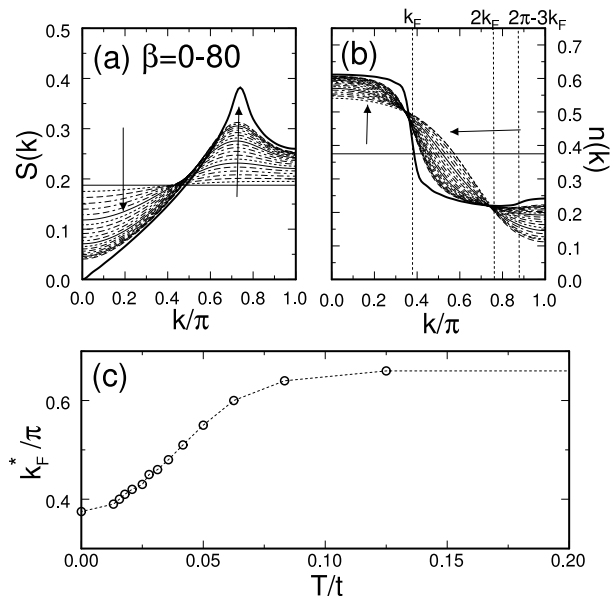,width=80mm,angle=00}
\caption{ 
(a) Spin structure factor, and (b) momentum distribution of a $t-J$ chain of length $L=32$ and $N=24$ fermions, with $J=0.05$, for different values of the temperature. The thick full lines in (a) and (b) correspond to $T=0$. 
Arrows indicate increasing $\beta$ (decreasing temperature) in units of $1/t$, in steps of four.
(c) shows the behavior of the ``Fermi momentum" $k_F^*$ as a function of temperature, obtained as the inflection point in the momentum distributions shown (b). 
} 
\label{fig3}
\end{figure}
\end{centering}

At $\beta=0$, the state $|\psi(0)\rangle$ is the maximally entangled state between
the real system and the fictitious system. We can see that this is independent of 
the representation, and we can choose any arbitrary basis.
 In particular, is natural to work in an occupation number representation
where the state of each site $i$ takes on a
definite value $n_i$. One finds
\begin{equation}
| \psi(0) \rangle = \prod_i \sum_{n_i} |n_i \tilde{n}_i\rangle = \prod_i
|I_{0i}\rangle \label{site}, \,
\end{equation}
defining the maximally entangled state $|I_{0i}\rangle$ of site $i$ with
its ``ancilla'', the local degree of freedom in the auxiliary system.
At this point it becomes convenient to perform a time-reversal transformation on the ancillas. Therefore, for the case that concerns us, where double-occupancy is forbidden, this state can be written as $|I_{0i}\rangle=|\uparrow,\tilde{\downarrow}\rangle - |\downarrow,\tilde{\uparrow}\rangle + |0,\tilde{0}\rangle$. This simple step allows us to work in a basis where the total spin projection $S_{\rm{tot}}^z$ of the chain-ancilla system is effectively zero.\cite{Feiguin2005a}  We emphasize that both spin and charge degrees of freedom appear in $|I_{0i}\rangle$, and are therefore treated on equal footing as regards finite-temperature effects. (Subject, of course, to the no double occupancy constraint.)

The state of the system at an arbitrary temperature $\beta$ is obtained by 
evolving the maximally mixed state in imaginary time, Eq.~(\ref{thermoa}) with $\beta=0$, using 
the Hamiltonian acting on the real degrees of freedom. The ancillas do not have 
any interactions controlling their dynamics.  They evolve only by their 
entanglement with the physical spins, effectively acting as a thermal bath. This 
is the basis of the finite-temperature DMRG method.\cite{Feiguin2005a} Notice 
that at zero temperature, the site and the ancilla are totally disentangled, 
while at finite temperature there is always a finite degree of entanglement that 
only depends on the dynamics of the system.

An important consequence of the previous description  
is that it would correspond to working in the grand canonical ensemble: Even though the spin and charge quantum numbers are conserved for the enlarged system, this is not the case if we restrict ourselves to the physical chain. In order to work in the canonical ensemble we need to start from a thermal vacuum where the physical states $|n\rangle$ and their copies $|\tilde{n}\rangle$ have each a fixed number of particles. 
To achieve that, we are going to construct a state that is a sum of all possible states of charge and spin, with the constraint that the total number of particles on the chain has to be equal to $N$, and that the charge state of the ancillas is an exact copy of the charge state of the physical chain. We achieve this by calculating the ground state of a very peculiar Hamiltonian, using conventional DMRG:
\begin{eqnarray}
H & = & -\sum\limits_{i\neq j} \left(\Delta^\dagger_{i} \Delta_{j}+\mathrm{h.c.}\right).
\label{entangler}
\end{eqnarray}
The operator $\Delta^{\dagger}$($\Delta$) creates (annihilates) a singlet between the physical spin and the ancilla,
\begin{equation}
\Delta^\dagger_i = \left(c^\dagger_{i\uparrow}\tilde c^\dagger_{i\downarrow}-c^\dagger_{i\downarrow} \tilde c^\dagger_{i\uparrow} \right) / \sqrt{2},
\end{equation}
where the ``tilde'' operators act on the ancillas on site $i$. 
The ground state of this Hamiltonian is precisely the equal superposition of all the configurations of $N$ ``physical site-ancilla'' singlets on $L$ sites.
This state can be represented very efficiently in terms of a matrix-product state, and consequently, by the DMRG method. In practice, the number of DMRG states required is of the order of the number of particles.
We find the use of the ``entangler'' Hamiltonian practical and convenient. Note that it does not disrupt the SU(2) symmetry of the $t-J$ model. 

\section{Green's functions}
We study the spectral properties of a spin-incoherent chain by evaluating 
the Green's functions at time $t$ at finite spin temperature:
\begin{equation}
G(x-x_0,t,\beta)=\langle \sill(\beta)|e^{iH_{t-J}t}\hat O^{\dagger}(x)e^{-iH_{t-J}t}\hat O(x_0)|\sill(\beta)\rangle,
\label{gf}
\end{equation}
where the generic operators of interest $\hat O(x)$,$\hat O^\dagger(x)$ act on the system at site $x$, and the Hamiltonian $H_{t-J}$ governs the physics of the actual physical chain, not including the ancillas.

We use a similar method to the one described in Ref.[\onlinecite{Barthel2009}]. The calculation proceeds as follows: First, we evolve the maximally entangled state in imaginary time to the desired value of $\beta$ measured in units of $1/t$, {\it e.g.} $\beta=2$ means $T=t/2$. Then, an operator $\hat O(x_0=L/2)$ is applied in the center of the chain. The resulting state is evolved in real-time, and at every step
we measure the overlap with the state $\hat O(x)e^{-iH_{t-J}t}|\sill(\beta)\rangle$.
We obtain the desired Green's function in frequency and momentum by Fourier transforming the results in real space and time.
Both states, $|\sill(\beta)\rangle$, and $\hat O(x_0)|\sill(\beta)\rangle$, have to be evolved in real time. In this work we use a third order Suzuki-Trotter decomposition with a typical time-step $\tau=0.1$, both for the real-time and imaginary-time parts of the simulation, keeping 800 states, enough to maintain the truncation error below $10^{-5}$. 
As customary in most DMRG calculations, we used open boundary conditions, and by doing the Fourier transform we are assuming that boundary effects can be ignored, as though the system were translational invariant.  In order to minimize the finite-size effects induced by the boundaries 
\cite{Eggert1996b,Mattsson1997,Kakashvili2006} 
we evolve to times $t=15$, and Fourier transform to frequency using a Gaussian window or width $\sigma = 6$ in the time domain, which in turn leads to a mode getting an artificial broadening in frequency proportional to $1/\sigma$. We point out that we have not used the linear prediction method introduced in Ref.[\onlinecite{Barthel2009}], but the bare data obtained from the simulation.
At zero temperature we have found that this works well, reproducing the features observed in the Bethe Ansatz solution of the Hubbard chain (compare our Fig.2(c) to Fig.7 in Ref.[\onlinecite{Ogata1990}]), namely, the singularity at $3k_F$ (seen at $2\pi-3k_F$). At finite temperatures the system develops a finite correlation length, which is further enhanced at higher temperatures due to the spin-incoherent mechanism -- to be discussed below. This is reflected in a localization that makes the boundary effects irrelevant. Working with open boundary conditions also avoids the degeneracy occurring in systems with periodic boundary conditions and size $L=4n$, with $n$ being an integer.\cite{Ogata1990}
The numerical errors can be attributed to the accumulation of truncation error, and the Trotter decomposition. The latter are under control, while the truncation error would translate into error bars that are much smaller than the broadening in frequency, and are therefore ignored for visualization purposes.

\section{Results}
In Figs.\ref{fig2} and \ref{fig3} we show some characteristic physical quantities at finite temperature, such as the specific heat $C_V$, spin structure factor $S(k)$, and momentum distribution function $n(k)$, for a chain with $L=32$ sites and $N=24$ fermions (3/8-filling). All results correspond to a value of $J=0.05$. The correlation functions are defined as:
\begin{equation}
S(k) = \frac{1}{L}\sum_{i,j} \langle S_i^z S_j^z\rangle e^{ik(i-j)} \,,
\end{equation}
\begin{equation}
n(k) = \frac{1}{L}\sum_{i,j} \langle c_{i\uparrow}^\dagger c_{j\uparrow}\rangle e^{ik(i-j)}\;.
\end{equation} 
The specific heat in Fig.\ref{fig2} shows a clear peak at a value of the temperature $T \sim J$, signaling the onset of the spin-incoherent regime where the spin degrees of freedom are highly thermalized.  At larger temperatures, the spin degrees of freedom are saturated, but a broad peak associated with the charge degrees of freedom is apparent. The spin structure factor in Fig.\ref{fig3}(a) shows a peak at momentum $k=2k_F$ that develops precisely at low temperatures $T \lesssim J$. The zero temperature result is consistent with that computed in Ref.[\onlinecite{Ogata1990}].
The momentum distribution in Fig.\ref{fig3}(b,c) displays the expected Luttinger liquid profile, with no discontinuities at the Fermi point. 
Below this value of the temperature we also notice the onset of a singularity at $k=3k_F$ in the momentum distribution.\cite{Matveev2007} This singularity corresponds to the transfer of spectral weight to the shadow bands, that originate from the scattering with the spin fluctuations that diverge at $k=2k_F$.\cite{Ogata1990,Penc1996,Favand1997}
While this behavior had already been seen in finite-temperature calculations using the factorized wave function (\ref{gs}) in Ref.[\onlinecite{Penc1997}], our calculations do not rely on the factorized wavefunction, or the XY approximation in the spin sector (rather than full Heisenberg symmetry).  Moreover, they can be readily generalized to a number of other spin symmetries, including the incorporation of spin-orbit effects.

\begin{centering}
\begin{figure}
\epsfig {file=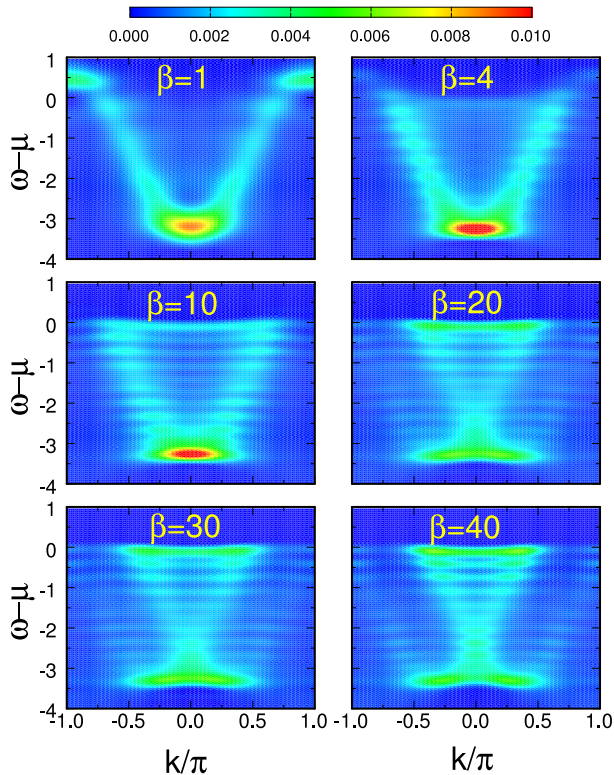,width=80mm,angle=00}
\caption{Photoemission spectrum of a one-dimensional $t-J$ chain with $L=32$ sites and $N=24$ fermions, and $J=0.05$. Different panels show different values of temperature $T=1/\beta$. The crossover to the spin-incoherent regime is achieved at $\beta \simeq 20$.
} 
\label{fig4}
\end{figure}
\end{centering}

We approximated the {\it temperature-dependent} Fermi momentum $k_F^*$ by taking the inflection point where $n(k)$ changes curvature, and plotted the result in Fig.\ref{fig3}(c). The Fermi momentum moves continuously from the zero-temperature value $k_F=\pi N/2L$ to $2k_F$, with the crossover region centered around $T \sim J$, as expected.\cite{Cheianov2005} We actually observe a saturation value below $2k_F$, but this is an artifact of taking the non-rigorous definition of $k_F$ as the inflection point in $n(k)$. It is important to note that $n(k)$ for the Hubbard or $t-J$ models changes its form qualitatively for fillings larger than 1/4, but less than 1/2.\cite{Ogata1990}  

The ``special" value of 1/4 filling in the lattice models (as opposed to the effectively low-energy theories) is related to the underlying Hubbard model, and the qualitative change for fillings above it and below it can be understood in the following way:
The shadow bands carry a significant ammount of spectral weight above $k=k_F$. When the density is 
larger than quarter-filling, the shadow band covers all $k$-space 
 from 0 to $\pi$. This translates into and increase of weight above $k_F$. 
(Recall that $n(k)$ is the integrated weight for $\omega<0$.\cite{Penc1996,Favand1997,in progress}) 

One further feature of $n(k)$ is particularly striking: The values $n(k_F)$ and $n(2k_F)$ are temperature independent within the accuracy of our calculation, with $k_F\equiv \pi N/2L$ and $2k_F$ twice $k_F$, rather than the value obtained by the inflection point of $n(k)$.   The same behavior was also observed in the factorized wavefunction approach with XY symmetry described in Ref.[\onlinecite{Penc1997}].  Evidently, then, it is independent of the spin symmetry, and probably results from an {\em effective} temperature independence  of the charge sector (temperature is taken to be precisely zero with respect to the charge sector in Ref.[\onlinecite{Penc1997}]). Since both calculations are effectively in the large $U$ limit, it may be that the temperature independence of these two points can be attributed to two extreme ``charge configurations"--one ``evenly spaced" and one "maximally paired" (two sitting right next to each other).  In the density-density correlations, the former would correspond to $4k_F$ oscillations and the latter to $2k_F$ oscillations.  In the large $U$ limit, it must be that these are both extreme ``spin-independent" configurations in the sense that  ``evenly spaced" electrons or ``maximally paired" paired electrons have only minimal contributions from the spin energy, leading to the temperature independence of $n(k_F)$ and $n(2k_F)$. 

Qualitatively, the shift from $k_F$ to $2k_F$ (as measured by the inflection point of $n(k)$) when the spin-incoherent regime is obtained can be understood as a shift from particles with spin dynamics to particles that are effectively spinless.\cite{Cheianov2005,Fiete2004} In the large but finite $U$ limit of the Hubbard model, electrons at zero temperature ``dimerize" ever so slightly and in this way maintain a ``memory" of their non-interacting $k_F$.  However, once $T\sim J$, this dimerization is washed out (because the energy scale for dimerization is set by $J$) and  effectively shifts $k_F$ to its ``spinless" value, $2k_F$.\cite{Fiete2007b}

Fig.\ref{fig4} shows the momentum resolved photoemission spectrum obtained by taking $\hat O(x)=c_\uparrow(x)$ in Eq.(\ref{gf}) in the previous treatment.
At infinite temperature, it resembles a band of non-interacting spinless fermions, following a $-2t\cos(k)$ dispersion with a ``width" much larger than seen in a zero temperature calculation ({\it e.g.}, the result in Fig.\ref{fig1} for a larger system size).  As the temperature is lowered, and $\beta$ increased, we see spectral weight being transfered from positive to negative energies.  At the same time, the band appears to broaden in the momentum direction, also splitting into seemingly discrete weakly dispersive levels. At a value of $\beta \sim 10$, the strongest features of the dispersion describe a number of discrete levels on top of a band that follows the spinless dispersion, with very small weight above the Fermi level.  The high weight at zero momentum corresponds to the high density of states in the van Hove singularity of the spinless band. At temperatures below $T=1/20=J$, the dispersion splits into two ``echoes", centered at $k=\pm k_F$, showing the emergence of the shadow bands and features more reminiscent of a LL.  We have verified that the gap between the horizontal levels are a finite-size effect, and the spacing grows as $1/L$ as we reduce the system size. The two-peaked features for $\beta \leq 10$ in the horizontal dispersion correspond to scattering of charge states with the non-dispersive spins present in the spin-incoherent regime. 

\begin{centering}
\begin{figure}
\epsfig {file=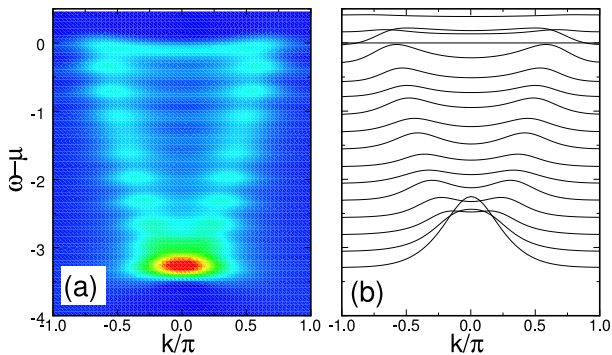,width=80mm,angle=-00}
\caption{ 
Photomoemission results for the same system as Fig.\ref{fig4}, and $\beta=10$. The right panel shows the same data with the amplitude represented as lines.  Note the qualitative agreement with the experimental results of Ref.[\onlinecite{Steinberg2006}].  This indicates that disorder effects are not needed to explain the main features of the data, and spin-incoherent physics is likely relevant.}
\label{fig5}
\end{figure}
\end{centering}

Many of these features can be qualitatively understood within SILL theory.  First imagine a system at zero temperature with $J \ll t$.  At the fillings we consider, this system will behave as a LL because it is gapless.  The spectral function will exhibit cusp-like singularities at $\omega=v_\sigma k$ and $\omega=v_\rho k$ where $v_\sigma$ is the spin velocity and $v_\rho$ is the charge velocity. \cite{Voit1993,Meden1992}  Since $J\ll t$, $v_\sigma \ll v_\rho$. The ``sharpness" of the cusps are determined by the interaction parameters of the LL theory in the spin and charge sectors.\cite{GiamarchiBook,GogolinBook}  If one now adds a small finite temperature (so as to remain in the LL regime) the LL correlation functions obtain a finite correlation length $\xi \propto v_\sigma/T$.  This correlation length will smear and broaden the cusp-like singularities. \cite{Iucci2007}  As the temperature is further raised, there is a smallest correlation length than can be obtained: the interparticle spacing.  In the spin-incoherent regime, $\xi$ effectively saturates at this value and leads to a universal broadening\cite{Fiete2004,Cheianov2004,Fiete2005} of $\sim \ln(2)k_F/\pi$  of the singularity associated with the charge mode and a vanishing of the singularity associated with the spin mode.  This effect is evident in Fig.\ref{fig4} for $\beta=1$ when one compares to the zero temperature result in Fig.\ref{fig1}.  For $\beta \sim 1/J$ the shadow bands are beginning to emerge and the spin degrees of freedom are starting to become dynamical leading to a complicated spectral form.

Finally, we focus on the crossover regime at $\beta=10$. The spectrum is shown with clarity in Fig.\ref{fig5}. These results can be compared to experiments in nanowires.\cite{Steinberg2006} (See {\it e.g.}, their Fig.3.) The qualitative agreement indicates that the experiments were most likely in the regime of highly thermalized spin states. Moreover, our calculations conclusively demonstrate that disorder effects need not be invoked to explain the data.\cite{Steinberg2006,Halperin2007}  These results can also be compared to high-energy angle-resolved photoemission experiments on quasi one-dimensional SrCuO$_2$, where the V-shaped dispersion is also observed.\cite{Suga2004} 

\section{Summary and Conclusions}

We have presented a numerical study of the spectral properties of $t-J$ chains at finite temperature, using a generalization of time-dependent DMRG techniques that combines evolution in real and imaginary time.  The study of finite temperature effects on the spectral functions of one-dimensional systems using quantum Monte Carlo techniques \cite{Suga2004,Matsueda2005,Abendschein2006} has mostly focused on the interpretation of photoemission experiments on quasi one-dimensional compounds such as SrCuO$_2$\cite{Kim1996,Kim1997,Kim2006} and TTF-TCNQ.\cite{Sing2003} While the Monte Carlo technique is free of the sign problem in one-dimension, the calculation of spectral properties involves an analytic continuation from Matsubara frequencies which is not straightforward in the spin-incoherent regime. The application of a maximum entropy method to the results is affected by statistical uncertainties, inherent from the stochastic QMC approach.  On the other hand, our method is naturally applied to study the spin-incoherent regime and we have demonstrated it is quantitatively accurate by comparison with Bethe ansatz results in various limits. 

We have clearly seen that at temperatures of the order of $T \sim J$, the system experiences a crossover from a spin-coherent to a spin-incoherent regime, which is clearly manifest in the spectra. 
Our results in finite systems show a compelling qualitative agreement with experiments in nanowires.\cite{Steinberg2006} The fact that our systems have a finite size works to our advantage since our parameters are similar to the experimental conditions, which involve wires at low densities with few electrons. 

In summary, we have been able to address an important and analytically inaccessible regime of strongly correlated one-dimensional systems. The time-dependent DMRG method has the power to access the full crossover from SILL to LL behavior as a function of temperature and therefore is a powerful tool in the interpretation of experimental results, and as a guide to analytical approximations not yet developed.   The technique can be readily adapted to study a number of related problems, including cold atomic gases which are notoriously plagued by finite temperature effects. \cite{Kakashvili2008}

\section*{Acknowledgements}

GAF gratefully acknowledges support the Lee A. DuBridge Foundation and ARO grant W911NF-09-1-0527. AEF would like to thank M. Troyer, M. Hastings, and A. Yacoby for useful discussions.




\end{document}